\documentstyle[12pt,graphicx]{article}
\title{ Analytic Spectra of CMB Anisotropies and Polarization
Generated by Relic Gravitational Waves with Modification due to
Neutrino Free-Streaming }
\author{T.Y. Xia  and  Y. Zhang \thanks{yzh@ustc.edu.cn} \\
        Astrophysics Center \\
        University of Science and Technology of China \\
        Hefei, Anhui, China }
 \date{}

\begin{document}
\maketitle
\baselineskip=21truept

\newcommand{\be}{\begin{equation}}
\newcommand{\ee}{\end{equation}}
\newcommand{\ba}{\begin{eqnarray}}
\newcommand{\ea}{\end{eqnarray}}

\sf
\begin{center}
\Large  Abstract
\end{center}
\begin{quote}
 {\large
We present an analytical calculation of the spectra of CMB anisotropies
and polarizations generated by relic gravitational waves (RGWs).
As a substantial extension to the previous studies,
three new ingredients  are included in this work.
Firstly,  the analytic $C_l^{TT}$ and  $C_l^{TE}$ are given;
especially the latter can be useful to extract signal of RGWs from
the observed data in the zero multipole method.
Secondly, a fitting formula of the decaying factor on small scales
is given,
coming from the visibility function around the photon decoupling.
Thirdly,
the impacts by the neutrino free-streaming (NFS) is examined,
a process that occurred in the early universe
and leaves observable imprints on CMB via RGWs.

It is found that the analytic $C_l^{TT}$ and $C_l^{TE}$
have profiles agreeing with the numeric ones,
except that $C^{TT}_l$ in a range $l \le 10$  and
the $1^{st}$ trough of  $C_l^{TE}$ around $l \sim 75$ have some deviations.
With the new damping factor,
the  analytic $C^{EE}_l$ and $C^{BB}_l$
match with the numeric ones with the maximum errors only $\sim 3\%$
up to the first three peaks  for $l\le 600$,
improving the previous studies substantially.
The correspondence of the positions of peaks of $C^{XX}_l$
and those of RGWs are also demonstrated explicitly.
We also find that NFS  reduces the amplitudes of  $C^{XX}_l$
 by $(20\% \sim 35\%)$ for $l\simeq(100\sim 600)$
and shifts slightly their  peaks to smaller angles.
Detailed analyses show that
the zero multipoles $l_0$, where $C_l^{TE}$ crosses $0$,
are shifted to larger values by NFS.
This shifting effect is as important as those caused
by different inflation models and different baryon fractions.
}
\end{quote}

PACS numbers: 98.70.Vc, 04.30.Nk, 98.80.-k,  95.85.Ry

Key words:
cosmic microwave background radiation,
relic gravitational waves,
polarizations,
cosmic neutrinos

\baselineskip=19truept

\begin{center}
{\Large 1. Introduction}
\end{center}

The observations on CMB
\cite{Spergel03,Peiris,spergel,Page,dunkley,hinshaw,komatsu,nolta}
are in good agreement with
a spatially flat universe
with nearly scale-invariant spectrum of primordial adiabatic
perturbations predicted by the inflation model.
Generally, two kinds of perturbations of the spacetime
metric are of interest:
density perturbations, i.e. scalar type \cite{Sasaki,HU02}
and relic gravitational wave (RGW), i.e. tensorial type
\cite{Rubakov:1982,Fabbri:1983,Abbott:1984,Starobinskii:1985,Muhhanov,
Basko,Polnarev,Zhao06a}, respectively.
Both perturbations will influence the CMB
anisotropies and polarizations
through the  Boltzmann equation for photons.
Although the contribution
from density perturbation is dominant,
RGWs may have important contributions \cite{komatsu}.
In particular, as a  special feature,
RGWs can give rise to magnetic type of CMB polarizations,
and this could provide a distinguished way
to directly detect RGWs of very long wavelength
comparable to the  Hubble radius $\sim 1/H_0$.
In comparison,
the usual laser interferometers, such as  LIGO,
probe the intermediate frequency
range $\nu = 50-1000$Hz \cite{LIGO2008},
the waveguide detectors probe the high frequency range
$\nu= 10^5- 10^7$Hz \cite{Cruise,Tong},
and the Gaussian laser beam detectors
probe in very high frequencies $\nu \sim 10^{10}$Hz
\cite{Li2000,Tong2008}.

The spectra of CMB anisotropies and polarizations generated by RGWs
have long been computed  \cite{Seljak,Lewis,
Hu95a,Zaldarriaga,Grishchuk,Kamionkowski}.
In particular, by approximate treatments of the photon decoupling,
Refs.\cite{Pritchard,Zhao06}
have derived  the analytic expressions of the polarization  spectra,
$C_l^{EE}$ and $C_l^{BB}$,
which show explicitly the influences of RGWs,
the inflation, the decoupling process,
the baryons, and the dark energy, etc.
But, compared with the numeric computations,
these  analytic $C_l^{EE}$ and $C_l^{BB}$
have large errors and are valid only in a limited region of $l\le 300$.
This is largely due to the damping factor $D(k)$
coming from the visibility function during the decoupling
is not accurate enough on small scales.
Thus, improvements of accuracy and extensions to a broader range
are certainly desired.
We will present a fitting formula of $D(k)$,
which substantially improves both the accuracy and the region of validity
over the previous studies.

Moreover, in the previous analytic calculations \cite{Pritchard,Zhao06},
the cross spectrum $C_l^{TE}$  was not given,
neither was the temperature spectrum $C_l^{TT}$.
Theoretically, the magnetic type
of polarization $C_l^{BB}$ can only be generated by RGWs.
But the current observed data of $C_l^{BB}$
is not yet sufficient to confirm
the existence of RGWs \cite{Page,dunkley,komatsu,hinshaw,nolta}.
On the other hand,
the cross spectrum $C_l^{TE}$ is about two orders higher than
$C_l^{BB}$, and also contains the contribution from RGWs
beside the density perturbations.
More importantly,
WMAP5 has detected $C_l^{TE}$ \cite{komatsu}.
This gives rise to  the possibility to extract RGWs,
since the contributions from the scalar and tensorial perturbations
behave differently.
In particular, RGWs can change the value of the multipole $l_0$,
where $C_l^{TE}$ first crosses $0$.
Thereby, one can, in principle, tell whether there is contribution from RGWs,
and give constraints on the scalar/tentor  ratio
and other cosmological parameters.
This so-called zero multipole method  relies on
detailed analysis of $C_l^{TE}$  \cite{baskaran,Polnarev08}.
Since $C_l^{TE}$ depends on several cosmological parameters,
analytic results are always helpful in exhibiting their properties.
In this paper we give the analytic $C_l^{TE}$,  as well as $C_l^{TT}$,
due to RGWs.

As a source to CMB,
the RGWs depends on the  inflation and on the dark energy \cite{zhang06b}.
Besides, it also depends on  physical processes
in the radiation-dominated Universe,
such as NFS \cite{Weinberg,Dicus,Watanabe,Miao07},
the QCD transition,  and the $e^{\pm}$ annihilation
\cite{Watanabe,wangzhang08,schwarz}.
While the latter two processes are effective
only on small scales $\nu>10^{-12}$ Hz
and do not appear in the currently observed  CMB spectra,
the former process is effective on  large scales
with a frequency region $\nu \simeq (10^{-17}\sim 10^{-10})$ Hz,
reducing the amplitude of RGWs by $\sim 20\%$ \cite{Miao07}.
This in turn will have observable effects on
the $2^{nd}$ and $3^{rd}$ peaks of  $C_l^{XX}$.
Thus,  we will
employ the RGWs spectrum modified by NFS  \cite{Miao07}
to calculate  $C_l^{XX}$,
 improving the previous calculation in Ref.\cite{Zhao06}
that did not consider the effect of NFS.

In Section 2 we review briefly the result of RGWs spectrum $h(\nu,\eta)$
with modifications due to NFS.
In Section 3 we will use this $h(\nu,\eta)$ to compute
the spectra $C_l^{TT}$, $C_l^{TE}$, $C_l^{EE}$, and $C_l^{BB}$.
The Basko-Polnarev's method will be used
\cite{Basko,Polnarev,Baskaran,Keating}.
In the process of the time integration,
a fitting formula of damping factor $D(k)$ on small scales will introduced,
which gives a better representation of
the visibility function $V(\eta)$ during the photon decoupling.
Section 4 examines the influences on $C_l^{XX}$
due to NFS, the spectrum index of inflation,  and the fraction of baryon,
especially, the corresponding shifting of $l_0$ of $C_l^{TE}$
is investigated.
A summary is given in Section 5.
We use the unit in which $c=\hbar =k_B= 1$ in this paper.

\begin{center}
{\Large 2.  RGWs Modified by NFS}
\end{center}

The expansion of a spatially flat ($\Omega_\Lambda +\Omega_m+\Omega_r=1$)
Universe can be described by the Robertson-Walker  metric
\be
ds^2=a^2(\eta)[-d\eta^2+(\delta_{ij}+h_{ij})dx^idx^j],
\ee
where $\eta$ is the conformal time and
the small perturbation $h_{ij}$ is RGWs  and is taken to
be  traceless and transverse (TT gauge)
\be
h^i_{\,\,i}=0,\,\,\, \hspace{.2cm} h^{ij}_{\,\,\,,j}=0.
\ee
The wave equation of RGWs is
\be \label{eq1}
 \partial_{\nu}( \sqrt{-g}\partial^{\nu} h_{ij} )=0.
\ee
By the Fourier decomposition
\be \label{hij}
h_{ij} (\eta,{\bf x})=
\sum_{\sigma}\int\frac{d^3k}{(2\pi)^3}
\epsilon^{\sigma}_{ij}h_k^{(\sigma)}(\eta)
e^{i\bf{k}\cdot{x}}
\ee
for each mode $\bf k$ and
each polarization $\sigma =( +, \times)$,
Eq.(\ref{eq1}) can be put into the form
\be \label{heq}
\ddot{h}_k+2\frac{\dot{a}}{a}\dot{h}_k+k^2h_k=0,
\ee
where  $\dot h_k=d h_k/d\eta$,
the polarization index $\sigma$ has been skipped for simplicity.
Eq.(\ref{heq}) holds for most of the  stages
from the inflationary to the current accelerating expansion.
The explicitly form of the scale factor $a(\eta)$
 are given
\be \label{r}
a(\eta)=a_e(\eta-\eta_e),\,\,\,\,\eta_s\leq \eta\leq \eta_2
\ee
for the radiation-dominant stage,
\be \label{m}
a(\eta)=a_m(\eta-\eta_m)^2,\,\,\,\,\eta_2 \leq \eta\leq \eta_E
\ee
  for
the matter-dominant stage,
and
\be \label{accel}
a(\eta)=l_H|\eta-\eta_a|^{-\gamma},\,\,\,\,\eta_E \leq \eta\leq \eta_0
\ee
for the accelerating stage up to the present time $\eta_0$,
where  $\gamma \simeq 1.044$ for a dark energy $\Omega_{\Lambda}\simeq 0.75 $,
and $l_H=\gamma/H_0$ with $H_0$ being  the Hubble constant.
The normalization of $a(\eta)$ is such that $|\eta_0-\eta_a|=1$,
where $\eta_0 \sim 3.11$.
The notations in Eqs.(\ref{m}) and (\ref{accel})
are adopted from that in Ref.\cite{zhang06b,Miao07,wangzhang08}.
In our convention,
the conformal time $\eta$ is dimensionless,
the scale factor  $a(\eta)$ has the dimension of length.
The analytic solution of Eq.(\ref{heq}) and the spectrum
are obtained for the expanding universe with the consecutive stages:
inflationary, reheating,
radiation-dominant, matter-dominant,
and accelerating, respectively in Ref.\cite{zhang06b}.

As also evidenced by the five-year WMAP
\cite{dunkley,komatsu,hinshaw,nolta},
there exists a cosmic neutrino background
with the three light species.
By the standard scenario of Big-Bang, during the radiation stage,
from the temperature $T \simeq 2$ MeV up to the beginning
of the matter domination,
the neutrinos are decoupled from other components and
start to freely stream in space.
This neutrino free-streaming gives rise to an anisotropic portion
of the stress tensor, which serves a source for RGWs.
Consequently,  during this  period $\eta_{\nu d}<\eta< \eta_2$,
Eq.(\ref{heq}) is modified to the following differential-integral equation
\cite{Weinberg,Dicus,Watanabe,Miao07}
\ba \label{RGW}
\ddot{h}_k(\eta)+2\frac{\dot{a}}{a}\dot{h}_k(\eta)+k^2h_k(\eta)=
          -24f_{\nu}(\frac{\dot{a}}{a})^2
\int_{\eta_{\nu d} }^{\eta} \dot{h}_k(\eta') K(k(\eta-\eta')) d\eta',
\ea
where the kernel of the integral is
\be \label{kernel}
K(x)\equiv -\frac{\sin x}{x^3}-\frac{3\cos x}
{x^4}+\frac{3\sin x}{x^5},
\ee
and $f_{\nu}=\rho_{\nu}/\rho_0$ is the fractional energy density
of neutrinos,
whose initial value is  $f_{\nu}(\eta=0)=0.40523$
for the effective number of species of neutrinos $N_\nu=3$.
The term on the right hand side of Eq.(\ref{RGW})
represents the anisotropic stress tensor due to NFS.

 Eq.(\ref{RGW}) has been solved by perturbations,
yielding the full analytic solution $h_k(\eta)$,
from the inflation up to the present accelerating stage \cite{Miao07},
and  it has been found  that
NFS causes a damping of $h_k$ by $\sim 20\%$
in the frequency range
\be \label{frequency}
\nu\simeq (10^{-17}, 10^{-10}) \,\,\, {\rm  Hz}.
\ee
Since  $\nu$ is related to  the conformal wavenumber $k$
 as  $\nu = \frac{H_0}{2\pi\gamma}k $
with $H_0=3.24\times 10^{-18}h$ Hz being the Hubble frequency,
Eq.(\ref{frequency}) corresponds to a range of the conformal wavenumber
\be \label{krange}
k \simeq (3\times 10^1,3\times 10^8)
\ee
for a Hubble parameter $h \simeq 0.7$.
NFS also slightly drags  the RGWs spectrum to small scales.
This dragging effect can be understood by a qualitatively analysis.
$K(x)$ in Eq.(\ref{kernel})
has a peak around $x\sim 0$ and $K(0)\sim 0.07$,
and its derivative can be roughly approximated as
$K'(x) \simeq -\delta(x)$.
The integration on the right hand side of Eq. (\ref{RGW})
can be integrated by parts
\be
\int^\eta_{\eta_{\nu d}} \dot{h}_k(\eta') K(\eta-\eta')  d\eta'
\simeq h_k(\eta)(K(0)-1).
\ee
Then Eq. (\ref{RGW}) is approximately reduced to
\ba \label{modifiedeq}
\ddot{h}_k+2\frac{\dot{a}}{a}\dot{h}_k+\left[ k^2- 24f_\nu (1-K(0))
\left( \frac{\dot{a}}{a}\right)^2\right]h_k =0.
\ea
By comparing Eq.(\ref{modifiedeq}) with Eq.(\ref{heq}),
one sees that  NFS modifies
the squared wave number $k^2$ to an effective one
\be \label{k'}
\bar k ^2 \equiv  k^2- 24f_\nu (1-K(0))\left( \frac{\dot{a}}{a}\right)^2<k^2.
\ee
If the mode $h_k(\eta)$ without NFS has a peak at $k=k_p$,
then the corresponding mode $h_k(\eta)$ with NFS will
have a peak at $\bar k(k)= k_p$,
which yields
$k\sim [1+12f_\nu (1-K(0))\left( \frac{\dot{a}}{a}\right)^2]k_p$.
The larger the $k_p$ is,
the greater the shifting amount is.
This analysis qualitatively explains
why  NFS slightly drags the peaks of RWGs to large $k$.
It is expected that
NFS will  cause
a slight shift of $C_l^{XX}$ to large $l$ via  RGWs consequently.

Since the damping  range of RGWs is $(10^{-17}, 10^{-10})$ Hz,
its lower frequency part just falls into the observable domain  of $C^{XX}_l$.
Therefore, in calculation of  CMB spectra,
the RGWs damped by NFS should be used as the source.
As will be seen in the next section,
the mode functions $h_k(\eta_d)$  and $\dot{h}_k(\eta_d)$ at the photon
decoupling time $\eta_d$, i.e., $z\sim 1100$,
will appear in the integral
expressions of the spectra of CMB anisotropies and polarizations.
They are plotted  in Fig.\ref{hhdot}.
The modifications on $h_k(\eta_d)$  and $\dot{h}_k(\eta_d)$
by NFS  leave observable imprints in the spectra of CMB.

As the initial condition,
the spectrum of RGWs
at the time $\eta_i$ of the horizon-crossing
during the inflation is chosen to be \cite{zhang06b,Miao07}
\be \label{initialspectrum}
h(\nu, \eta_i) =\frac{2k^{3/2}}{\pi}|h_k(\eta_i)|
=A(\frac{k}{k_H})^{2+\beta_{inf}},
\ee
where  $k_H = 2\pi$ is the comoving wave number and
corresponds to a physical wave $\lambda_H=2\pi a(\eta_0)/k_H=l_H$,
the constant $A$ is to be fixed by the observed CMB anisotropies,
and the spectrum index $\beta_{inf}$ is
a parameter determined by inflationary models.
The special case of $\beta_{inf}=-2$ is
the de Sitter expansion of inflation.
If the inflationary expansion is driven by a scalar field,
then the index $\beta_{inf}$ is related to
the so-called slow-roll parameters,
$\eta$ and $\epsilon$ \cite{Liddle},
as $\beta_{inf}=-2+(\eta-3\epsilon)$.
For demonstration purpose in our context,
we allow the parameter $\beta_{inf}$ to take the values $>-2$.
In literature, the RGWs spectrum is often written
in terms of $\Delta^2_h(k)$, related to Eq.(\ref{initialspectrum}) by
$h^2(\nu, \eta_i) =8 \Delta^2_h(k)$.
Without the running index, it is usually assumed to have
the form \cite{Spergel03} \cite{Peiris} \cite{Verde}
\be
\label{Delta}
\Delta^2_h(k)=A_T(\frac{k}{k_0})^{n_T}.
\ee
Here the tensorial spectrum index   $n_T=2\beta_{inf}+4$,
$k_0$ is some comoving pivot wavenumber,
whose corresponding physical wavenumber  is
$k_0/a(\eta_H)= 0.002$ Mpc$^{-1}$  \cite{Page,Verde},
and the amplitude $A_T=2.95\times10^{-9}r A(k_0)$ with
$A(k_0)\sim 0.8$ as determined by the WMAP observations accordingly
\cite{Peiris},
$r$ being the tensor/scalar ratio.
In general,  $r$ is model-dependent,
and  frequency-dependent \cite{Zhao06,baskaran}.
The value of $r$ has long been
an important issue \cite{seljak2,Cooray,Smith,Linde,Mukhanov}.
In our treatment, for simplicity, $r$ is only taken as a
constant parameter for normalization of RGWs.
Currently, only observational constraints on  $r$ have been given.
The 1-Year WMAP gives $r<0.71$ \cite{Spergel03}.
The 3-Year WMAP constraint based on the CMB polarization
gives $r<2.2 $ ($95\%$ CL) evaluated at $k_0$ \cite{Page},
and the full WMAP constraint is $r<0.55 $ ($95\%$ CL) \cite{spergel,Page}.
Recently, the 5-year WMAP data improves the upper limit to
$r<0.43$ ($95\%$ CL) \cite{dunkley}, and combined with BAO and SN
gives $r<0.2$ ($95\%$ CL) \cite{komatsu} \cite{hinshaw}.
The combination from such observations, as of
the Lyman-$\alpha $ forest power spectrum from SDSS, 3-year WMAP,
supernovae SN, and galaxy clustering,
gives an upper limit $r<0.22$
($95\%$ CL) and $r< 0.37$ ($99.9\%$ CL) \cite{seljak1}.
For concreteness,
we take $r\simeq 0.37$ in our calculation.

\begin{center}
{\Large 3. Analytical Spectra Of CMB}
\end{center}

In the Basko-Polnarev's method \cite{Basko,Polnarev},
the Boltzmann equation of the CMB photon gas for the $k$-mode
is written as a set of two coupled differential equations
\be \label{eqxi}
\dot{\xi}_k+[ik\mu +q]\xi_k= \dot{h}_k,
\ee
\be \label{eqbeta}
\dot{\beta_k}+[ik\mu +q]\beta_k=qG_k.
\ee
where  $\beta_k$ represents the linear polarization,
$\alpha_k \equiv\xi_k-\beta_k$ represents
the anisotropy of radiation intensity, $\mu= \cos\theta$,
 $q$ is the differential optical depth,
and $G_k(\eta)=\frac{3}{16}
\int^1_{-1}d \mu'[(1+\mu'^2)^2\beta_k-\frac{1}{2}(1-\mu'^2)^2\xi_k]$
\cite{Zhao06}.
Note that the gravitational waves  $\dot{h}_k $ in Eq.(\ref{eqxi})
is the Sachs-Wolfe term \cite{SachsWolfe} and
plays the role of source to the temperature anisotropies.
In the following,
we omit the subscript $k$ for simplicity of notation.
The formal solutions of Eqs.(\ref{eqxi}) and (\ref{eqbeta})
can be written as:
\be \label{xi}
\xi(\eta,\mu)=\int^{\eta}_0 \dot{h}(\eta')
e^{-\kappa(\eta,\eta')}e^{ik\mu(\eta'-\eta)}d \eta',
\ee
\be \label{beta}
\beta(\eta,\mu)=\int^{\eta}_0 G(\eta')q(\eta')
e^{-\kappa(\eta,\eta')}e^{ik\mu(\eta'-\eta)}d \eta',
\ee
where
$\kappa(\eta',\eta)\equiv \int_\eta^{\eta'} qd\eta
=\kappa(\eta)-\kappa(\eta')$ with
$\kappa(\eta)\equiv\kappa(\eta_0,\eta)$
being the optical depth, such that
$q(\eta)=-d \kappa(\eta_0,\eta)/d\eta$.
To get rid of the angle dependence,
$\xi$ and $\beta$ are usually decomposed in terms of
the Legendre components
\be\label{xill}
\xi_l(\eta) = \frac{1}{2}\int_{-1}^1 \,d\mu\,
\xi(\eta,\mu)P_l(\mu),
\ee
\be\label{betall}
\beta_l(\eta) =\frac{1}{2}\int_{-1}^1\, d\mu\,
\beta(\eta,\mu)P_l(\mu),
\ee
where $P_l$ is the Legendre functions.
Using the  expansion formula
\[
 e^{ix\mu}
      =\sum_{l=0 }^{\infty} (2l+1)i^lj_l(x)P_l(\mu)
\]
and the ortho-normal relation for Legendre functions,
one obtains
\be \label{xigeneral}
\xi_l(\eta_0)=
i^l \int^{\eta_0}_{0}
    e^{-\kappa(\eta)}\dot{h}(\eta)j_l(k(\eta-\eta_0)) d\eta,
\ee
\be \label{betae}
\beta_l(\eta_0) =
  i^l \int_0^{\eta_0} G(\eta)V(\eta)\, j_l(k(\eta-\eta_0))d\eta,
\ee
both being evaluated  at the present time $\eta_0$,
where
\be \label{visibilitydef}
V(\eta)=q(\eta)e^{-\kappa(\eta)}
\ee
is the visibility function for the decoupling.
As is known, $V(\eta)$ is a narrow function
peaked around the decoupling time $\eta_d$ with a width $\Delta \eta_d$.
It  phenomenologically describes the details of the decoupling process
\cite{jones-wise, peebles68, Hu95a}.
Ref.\cite{Pritchard} uses a single gaussian
function to fit $V$ approximately.
In Ref. \cite{Zhao06}, as an improvement,
the following two pieces of half gaussian function are used
\be\label{halfgaussian1}
V(\eta)=
  \left\{
\begin{array}{ll}
V(\eta_d) \exp\left(-\frac{(\eta-\eta_d)^2}{2
\Delta\eta_{d1}^2}\right),        ~~~(\eta\leq\eta_d) ,   \\
V(\eta_d) \exp\left(-\frac{(\eta-\eta_d)^2}{2
\Delta\eta_{d2}^2}\right),         ~~~(\eta>\eta_d),
\end{array}
     \right.
\ee
where the decoupling time $\eta_d \simeq 0.0707$ corresponding to
the redshift $z_d \simeq 1100$,
$\Delta\eta_{d1}=0.00639$, $\Delta\eta_{d2}=0.0117$, and
$(\Delta\eta_{d1}+\Delta\eta_{d2})/2=\Delta\eta_{d}$
is the thickness of the decoupling.
In absence of reionization,
the coefficient $V(\eta_d)$ in Eq.(\ref{halfgaussian1})
will be determined by the normalization
\be \label{normv}
\int_0^{\eta_0} V(\eta)d \eta =1.
\ee
We have checked that the error between Eq.(\ref{halfgaussian1})
to the numerically fitted formula given in
Refs.\cite{jones-wise, Hu95a} is very small,  $\sim 3.9\%$
in the interval $\eta> \eta_d$.
Compared with the single gaussian function in Ref.\cite{Pritchard},
Eq.(\ref{halfgaussian1}) improves
the description of the visibility function by $\sim 10\%$
in accuracy.
Substituting Eq.(\ref{halfgaussian1}) into Eq.(\ref{betae}),
after some treatment of the integration
over the variable $\eta$  \cite{Zhao06},
the approximate analytic solution of $\beta_l$ without reionization
has been arrived up to the second order of a small $1/q^2$ in the
tight coupling limit,
\be \label{betal}
\beta_l(\eta_0)=\frac{1}{17}\ln\frac{20}{3} i^l \Delta \eta_d
D(k)\dot{h}(\eta_d) j_l(k(\eta_d-\eta_0)).
\ee
where  $\dot{h}(\eta_d) $ is the time derivative  of RGWs
at the decoupling, and
\be \label{D}
D(k)=\frac{1}{2}
[e^{-c(k\Delta\eta_{d1})^2}+e^{-c(k\Delta\eta_{d2})^2}]
\ee
is the Fourier transformation of $V(\eta)$ in Eqs.(\ref{halfgaussian1})
with the parameter $c$ taking values in $[0,2]$.
Formally,
the occurrence of the damping factor $D(k)$
is due to the $\eta$ integration of Eq.(\ref{betae})
of the form
$ \int_{-\infty}^{\infty} e^{-\eta^2}e^{i p \eta}d\eta
  =e^{-\frac{p^2}{4}}\int_{-\infty}^{\infty}  e^{- \eta^2}d\eta$
since the integrand factor $\dot{h}(\eta) j_l(k(\eta-\eta_0))$
contains a mixture of  $e^{ik\eta}$ and $e^{-ik\eta}$.
From view point of physics,
$D(k)$ is generically expected
\cite{Peebles80, Hu95a},
because  the photons diffuse through the baryons
around the decoupling and the fluctuations are severely damped
within the thickness of the surface of the last scattering.
Therefore, $D(k)$ is very sensitive the thickness $\Delta\eta_d$.
However,  as an approximation, the fitting formula Eq.(\ref{D})
is not accurate enough on the small scales and
will cause an over-damping of amplitudes of $C^{XX}_l$ for larger $l$.
This is because, in the afore-mentioned derivation of Eq.(\ref{D}),
other time-dependent factors in $h_k(\eta)$
have been taken as constants during the decoupling.
Besides, other processes important on small scales
were not taken into account \cite{Bardeen, Hu95a}.
To improve Eq.(\ref{D}),
we adopt the following simple fitting formula
\be \label{DD}
D(k)=\frac{1}{2}
[e^{-c(k\Delta\eta_{d1})^b}
     +e^{-c(k\Delta\eta_{d2})^b}]
\ee
where $b$ is a parameter.
It will be find that a good fit with $c\simeq 0.6$ and $b\simeq 0.85$,
comparing with the numerical results.
One may even effectively simplify Eq.(\ref{DD})
by the following
\be \label{D1}
D(k)= e^{-c(k\Delta\eta_{d})^b}.
\ee
We find that Eq.(\ref{DD}) and Eq.(\ref{D1})
yield the almost overlapping spectra $C_l^{XX}$,
and the error between them is only $\leq 1\%$.
In comparison with the numeric computations,
$C_l^{XX}$ generated by both Eqs.(\ref{DD}) and (\ref{D1})
are much more accurate than those by Eq.(\ref{D}).

To evaluate the temperature anisotropies spectrum $C_l^{TT}$,
one needs an analytic solution for $\alpha_l$.
The integrand in  (\ref{xigeneral})
contains a factor $e^{-\kappa(\eta)}$,
which can be treated approximately.
Since the visibility function $V(\eta)$
is a narrow function  and can be roughly
viewed as a Dirac delta function,
so by the relation $V(\eta)=d( e^{-\kappa(\eta_0,\eta)} )/d\eta $,
one can treat the factor $e^{-\kappa(\eta)}$ as the step function
\be \label{step}
e^{-\kappa(\eta)} \simeq
          \left\{
\begin{array}{ll}
0 &   \ \ \ \ (\eta<\eta_{d}), \\
1 &   \ \ \ \   (\eta_{d}< \eta<\eta_{0}).
\end{array}
\right.
\ee
Substituting Eq.(\ref{step}) into
Eq.(\ref{xigeneral}) yields
\be \label{lxi}
\xi_l(\eta_0)=
i^l \int^{\eta_0}_{\eta_d}  \dot{h}(\eta)j_l(k(\eta-\eta_0)) d\eta,
\ee
which can be integrated by parts,
\be \label{bypart}
\xi_l(\eta_0)=  -i^l h(\eta_d)j_l(k(\eta_d-\eta_0))+
      i^l \int^{\eta_0}_{\eta_d} d \eta h(\eta)
  \frac{d }{d \eta}j_l(k(\eta-\eta_0)) ,
\ee
where a term containing $h(\eta_0)$ from the upper limit at $\eta_0$
has been neglected
since the amplitude $h(\eta_0)$ is
about three orders smaller than that of $h(\eta_d)$.
The remaining integration term in
Eq.(\ref{bypart}) is small and can be neglected \cite{Baskaran},
since $h(\eta)$ is smaller during the late time $(\eta_d\sim\eta_0)$
and $\frac{d }{d \eta}j_l$ is oscillating functions.
Thus one has the following approximate, analytic solution
\be \label{xil}
\xi_l(\eta_0)=-i^l h(\eta_d)j_l(k(\eta_d-\eta_0)).
\ee
From  Eqs.(\ref{betal}) and (\ref{xil}) follows the temperature anisotropies:
\ba \label{alphal}
\alpha_l(\eta_0)
=  -i^l     j_l(k(\eta_d-\eta_0))
\left[h(\eta_d)+\frac{1}{17}\ln\frac{20}{3}
\Delta \eta_d D(k)\dot{h}(\eta_d)\right],
\ea
to which both $h(\eta_d)$ and  $\dot{h}(\eta_d)$ contribute.
As our calculation shows,
the contribution of $\dot{h}(\eta_d)$
is about two orders smaller than that of $h(\eta_d)$.

In terms of $\alpha_l$ and $\beta_l$,
one  calculates $C^{XX}_l$ caused by RGWs \cite{Kamionkowski}
straightforwardly.
The temperature anisotropies
\ba \label{ctt}
C^{TT}_l=\frac{1}{8\pi}\frac{(l+2)!}{(l-2)!}
\int k^2dk \left|
\frac{\alpha_{l-2}(\eta_0) }{(2l-1)(2l+1)}-
\frac{2\alpha_{l}(\eta_0)}{(2l-1)(2l+3)}
+\frac{\alpha_{l+2}(\eta_0)}{(2l+1)(2l+3)}\right|^2,
\ea
the electric type of polarization
\ba \label{cee}
C^{EE}_l= \frac{1}{16\pi}
   \int k^2dk \left|\frac{(l+1)(l+2)\beta_{l-2}(\eta_0)}{(2l-1)(2l+1)}+
\frac{6(l-1)(l+2)\beta_{l}(\eta_0)}{(2l-1)(2l+3)}+\frac{l(l-1)
   \beta_{l+2}(\eta_0)}{(2l+1)(2l+3)} \right|^2 ,
\ea
where the second term in the integrand has the coefficient $6(l-1)(l+2)$,
different from that in Ref.{\cite{Kamionkowski}},
the magnetic type of polarization
\ba \label{cbb}
C^{BB}_l = \frac{1}{16\pi}\int k^2dk
     \left|\frac{2(l+2)\beta_{l-1}(\eta_0)}{(2l+1)}
     +\frac{2(l-1)\beta_{l+1}(\eta_0)}{(2l+1)} \right|^2 ,
\ea
and the temperature-polarization cross correlation spectrum
\ba \label{cte}
C^{TE}_l&=&\sqrt{\frac{1}{8\pi}\frac{(l+2)!}{(l-2)!}}
\sqrt{\frac{1}{16\pi}} \int k^2dk
\left[  \frac{\alpha_{l-2}(\eta_0) }{(2l-1)(2l+1)}
        - \frac{2 \alpha_{l}(\eta_0)}{(2l-1)(2l+3)}
        +\frac{\alpha_{l+2}(\eta_0)}{(2l+1)(2l+3)} \right]
\nonumber\\ &&\times
\left[\frac{(l+1)(l+2)\beta_{l-2}(\eta_0)}{(2l-1)(2l+1)}+
\frac{6(l-1)(l+2)\beta_{l}(\eta_0)}{(2l-1)(2l+3)}+\frac{l(l-1)
   \beta_{l+2}(\eta_0)}{(2l+1)(2l+3)} \right] ,
\ea
where the second term in the integrand has the coefficient $6(l-1)(l+2)$,
different from that in Ref.{\cite{Kamionkowski}}.
Substituting  the explicit expressions $\alpha_l$ and $\beta_l$
of Eqs.(\ref{betal}) and (\ref{xil}) into the above
spectra, one finally has
\ba \label{ctt2}
C^{TT}_l = \frac{1}{8\pi}\frac{(l+2)!}{(l-2)!}\int k^2dk
P_{Tl}^2(k(\eta_d-\eta_0))\left|h(\eta_d)+
\frac{1}{17}\ln\frac{20}{3} \Delta \eta_d D(k)\dot{h}(\eta_d)\right|^2,
\ea
\ba \label{cee2}
C^{EE}_l = \frac{1}{16\pi}\left(\frac{1}{17}\ln\frac{20}{3}\right)^2
\int k^2dk  P_{El}^2(k(\eta_d-\eta_0))
\Delta \eta_d^2 D^2(k)\left|\dot{h}(\eta_d)\right|^2,
\ea
\ba \label{cbb2}
C^{BB}_l = \frac{1}{16\pi}\left(\frac{1}{17}\ln\frac{20}{3}\right)^2
\int k^2dk   P_{Bl}^2(k(\eta_d-\eta_0))
\Delta \eta_d^2 D^2(k)\left|\dot{h}(\eta_d)\right|^2,
\ea
\ba \label{cte2}
C^{TE}_l =&& \frac{1}{136\sqrt{2}\pi}
\ln\frac{20}{3}\sqrt{\frac{(l+2)!}{(l-2)!}}
\int k^2dk
         P_{Tl}(k(\eta_d-\eta_0))P_{El}(k(\eta_d-\eta_0))\nonumber\\
&&\frac{1}{2}\left\{\left[-h(\eta_d)-\frac{1}{17}\ln\frac{20}{3}
\Delta \eta_d D(k)\dot{h}(\eta_d)\right]
\dot{h}^*(\eta_d) \right. \nonumber\\
&& \left.+\left[-h^*(\eta_d)-\frac{1}{17}\ln\frac{20}{3}
\Delta \eta_d D(k)\dot{h}^*(\eta_d)\right]
\dot{h}(\eta_d)\right\}\Delta \eta_d D(k),
\ea
where the projection factors are defined as   \cite{Keating}:
\ba \label{pt}
P_{Tl}(x)=\frac{j_{l-2}(x)}{(2l-1)(2l+1)}+
         \frac{2j_{l}(x)}{(2l-1)(2l+3)}+\frac{j_{l+2}(x)}{(2l+1)(2l+3)}
         =\frac{j_l(x)}{x^2},
\ea
\ba  \label{pe}
P_{El}(x)
&&=\frac{(l+1)(l+2)}{(2l-1)(2l+1)}j_{l-2}(x)
    -\frac{6(l-1)(l+2)}{(2l-1)(2l+3)}j_{l}(x)+\frac{l(l-1)}
{(2l+1)(2l+3)}j_{l+2}(x) \nonumber \\
&&=-[2-\frac{l(l-1)}{x^2}]j_l(x)+\frac{2}{x}j_{l-1}(x),
\ea
\be \label{pb}
P_{Bl}(x)
=\frac{2(l+2)}{(2l+1)}j_{l-1}(x)
       -\frac{2(l-1)}{(2l+1)}j_{l+1}(x) \nonumber \\
=2j_{l-1}(x)-2\frac{l-1}{x}j_l(x).
\ee

With  $h(\eta_d)$ and  $\dot{h}(\eta_d)$
given from the last section
and $D(k)$ from Eq.(\ref{DD}),
we compute $C^{XX}_l$ and plot them in Fig. \ref{COMPARE},
where the following values of respective parameters are taken:
the inflationary index $\beta_{inf}=-2.02$,
the dark energy $\Omega_\Lambda =0.75$,
the baryon density $\Omega_b=0.045$,
the neutrino species $N_\nu=3$,
the tensor/scalar ratio  $r=0.37$,
$c=0.6 $, and $b= 0.85$.
For comparison, in Fig. \ref{COMPARE}
the numerical results from  CAMB \cite{Lewis} are also plotted.
It is seen that the analytic
$C_l^{EE}$ and $C_l^{BB}$ agree very well with the numerical ones
for the range $l \le 600$ covering the first three peaks,
and the error is only $\sim 2\%$.
Comparing with the previous analytic
evaluation in Refs.\cite{Pritchard,Zhao06},
our result not only extend the range of validity
from  $l\le 300$ to $l \le 600$,
but also improves accuracy substantially.

For the spectra $C_l^{TT}$ and $C_l^{TE}$,
the profiles of our analytic result
also agree with the numerical ones fairly well,
except that $C^{TT}_l$ in a range $l \le 10$
and the $1^{st}$ trough of  $C_l^{TE}$ around $l \sim 75$ have some deviations.
For the purpose of extracting RGWs,
more important is  $C_l^{TE}$,
whose amplitude at the $1^{st}$ trough has a maximum deviation  $\sim 20 \%$
from that of the numerical CAMB.
This is due to the approximation of
the temperature anisotropies $\xi_l(\eta_0)$ in Eq.(\ref{xil}),
which is not accurate enough for very large scales.

The  profiles of $C^{XX}_l$
are largely determined by those of  $h(\eta_d)$ and $\dot{h}(\eta_d)$,
especially,
the peaks and troughs of $C^{XX}_l$ correspond to those of RGWs.
The integrands for $C^{XX}_l$
in Eqs. (\ref{ctt2}), (\ref{cee2}), (\ref{cbb2}),
and (\ref{cte2}) contain the respective
projection factors,  $P_{Tl}$, $P_{El}$, $P_{Bl}$,
which  are made of the spherical  Bessel's  functions.
Since  $j_l(x)$ is rather sharply peaked around $x \simeq l$
for large $l$,
consequently,
the projection factors as functions of $k$ are  peaked around
\be \label{k-l}
k(\eta_0-\eta_d) \simeq k \eta_0\simeq l.
\ee
Therefore,  $C^{XX}_l$ as  integrations  over $k$
will receive  major  contributions from
the integration domain $k\sim l/\eta_0$ \cite{Zhao06}:
\be \label{ctt-appr}
C^{TT}_l \propto \left|h(\eta_d)\right|^2_{k \simeq l/\eta_0},
\ee
\be \label{cee-appr}
C_l^{EE}, \,  C_l^{BB}
\propto   \left|\dot{h}(\eta_d)\right|^2 _{k \simeq l/\eta_0} D^2(k),
\ee
\be \label{cte-appr}
C^{TE}_l \propto
    h(\eta_d) \dot{h}(\eta_d)_{k \simeq l/\eta_0} D(k).
\ee
By Eq.(\ref{ctt-appr}),
the locations of the peaks of $C_l^{TT}$ is mainly
determined by $|h(\eta_d)|^2$.
Indeed, the right panel in Fig. \ref{BETcomh} shows
that the peaks and troughs of
$C^{TT}_l$ correspond to those of $|h(\eta_d)|^2$.
Similarly, by Eq.(\ref{cee-appr}),
the locations of the peaks of $C_l^{EE}$ and $ C_l^{BB}$
correspond to those  of  $|\dot{h}(\eta_d)|^2$,
shown in the left panel in Fig. \ref{BETcomh}.
The similar correspondence, as revealed by Eq.(\ref{cte-appr}),
of the locations of the peaks of $C_l^{TE}$
to those of $h(\eta_d) \dot{h}(\eta_d)_{k} $
is also confirmed,
but  the graph is not presented though in order to save room.
Besides, $C_l^{EE}$, $ C_l^{BB}$ and $C_l^{TE}$ also depend on
the damping factor $D(k)$,
leading to  strong damping of
$C_l^{EE}$, $ C_l^{BB}$ and $C_l^{TE}$ at large $l$,
as shown in  Fig. \ref{COMPARE} and Fig. \ref{BETcomh}.
Overall,
our analytic formulation yields a good approximation
of $C^{XX}_l$
in comparison with the numerical results.

We use the data from 5-year WMAP \cite{komatsu,nolta}
to limit the B-mode polarization  $C^{BB}_l$ generated by RGWs
in Fig. \ref{WMAP5}.
It is seen that the current observational data
can only put a rather loose limit on RGWs.
For $r=0.37$ the amplitude of the predicted spectrum
is about 2 orders  below the upper limit by WMAP5.
Improvements on the limit, or possible direct detections of
$C^{BB}_l$
are expected from more sensitive polarization measurements
by upcoming experiments,
such as Clover \cite{Clover}, EBEX \cite{EBEX}, QUIET \cite{QUIET},
Spider \cite{Spider}, and Planck \cite{Planck}.

\begin{center}
{\Large 4. Influences  by NFS, Inflation, and Baryons}
\end{center}

{\large \bf  The  NFS}

Let us analyze the  effect of NFS on the spectra $C_l^{XX}$.
To demonstrate this,
the  spectra $C_l^{XX}$ with and without NFS are plotted in Fig. \ref{TEB}.
The  $l\le 100$ portion of the spectra are not much affected
by NFS,
only on the scales of $l>200$,
are the spectra modified effectively.
The reduction of amplitudes of  $C_l^{TT}$,  $C_l^{EE}$ and  $C_l^{BB}$
by NFS is noticeable only  starting from the second peak.
For instance, the third peak of $C_l^{TT}$ is reduced by  $\sim 25\%$
and the fourth peak by  $\sim 35\%$.
Similar modifications also occur in the spectra
$C^{EE}_l$, $C^{BB}_l$, and $C^{TE}_l$.
These features of modifications can be understood as follows.
As shown in Eq.(\ref{krange}),
the damping of RGWs  is effective for the conformal wavenumbers $k>30$,
which, by the relation in Eq.(\ref{k-l}),
means that only those portion with $l \geq 100$ of CMB spectra
will  be affected by NFS.
Besides,  Figs. \ref{TEB} also shows that
NFS causes
a slight shift of the peak locations of $C_l^{XX}$ to larger $l$,
a feature to be expected,
since NFS  shifts the peaks of
$h(\eta_d)$ and $ \dot{h}(\eta_d)$ slightly to large $k$,
the peaks of $C_l^{XX}$ will be accordingly
shifted to larger $l$ by Eqs.(\ref{k-l}) - (\ref{cte-appr}).
Given the current precision level of observations on CMB,
these small modifications caused by NFS
will be  difficult to detect at the moment.

As mentioned in the Introduction,
the cross spectrum  $C^{TE}_l$
can be useful in revealing the presence of RGWs in the zero-multipole method
\cite{baskaran,Polnarev08,Baskaran,Keating,Grishchuk07}.
The 5-year WMAP \cite{komatsu,nolta}
has given  the observed  $C^{TE}_l$,
which is  negative (anti-correlation) in a range  $l \simeq (50, 220)$.
Theoretically, it
is a combination of contributions of
the density perturbations and the RGWs as well.
To search for the evidence of  RGWs,
one needs to disentangle the contribution of RGWs from the total.
The inclusion of NFS into the calculation
will cause a shifting of the position of the peaks of $C_l^{TE}$
to larger values of $l$,
and $\Delta l$ tends to increase with $l$.
For instance,  Fig.\ref{TEB} shows that,
without NFS,
 $C^{TE}_l<0$  for $l\le 136$ and $C^{TE}_l>0$  for $l\simeq (137,179)$.
When NFS is included,
$C^{TE}_l<0  \hspace{.3cm} {\rm for } \,\,\, l\le 136$,
$C^{TE}_l>0 \hspace{.3cm} {\rm for } \,\,\, l\simeq (137, 183)$.
In this low $l$ region
the shifting due to NFS is small $\Delta l \le   4$.
But, in the large $l$ region  the shifting is large,
say, around $l\simeq 500$, it is $\Delta l\sim 10$.
This analysis tells us that
the zero multipole around $l_0\sim 50$ is not strongly affected by NFS.
However,
if we look at the the second zero multipole $l\sim 220$,
at which $C^{TE}_l$ crosses $0$ once again and turns positive,
the influence by NFS is rather strong, $\Delta l\sim 5$.
More accurate observations of $C^{TE}_l$
and detailed analysis are needed,
before a definite conclusion can be drawn on the existence of RGWs.

{\large \bf The inflation}

The CMB spectra generated by RGWs
depend very sensitively on the initial spectrum $h(\nu, \eta_i)$
during the inflationary stage.
For the power-law form of $h(\nu, \eta_i)$
in  Eq. (\ref{initialspectrum}),
$C_l^{XX}$ depend on
both the amplitude $A$ and the index $\beta_{inf}$.
Fig. \ref{TEBb18} shows $C_l^{XX}$
for the cases of $\beta_{inf}=-1.8$ and $-2.02$
with  NFS being taken into account.
A larger index $\beta_{inf}$ yields
higher amplitudes of  $C_l^{EE}$ and $C_l^{BB}$ in the whole range of $l$,
agreeing with the previous result \cite{zhang06b},
and higher amplitude of $C_l^{TT}$ for the range $l>20$.
In the zero multipole method,
one is more interested in  $C_l^{TE}$ in the narrow  range $l\simeq (40, 60)$,
in which the first zero multipole $l_0$ should appears.
Firstly, as Fig. \ref{TEBb18} shows,
$C_l^{TE}$  is negative in this range,
and moreover,
a larger index $\beta_{inf}(=-1.8)$ yields
a steeper, down slope of $C_l^{TE}$ of negative amplitude.
With other parameters being fixed,
a larger index $\beta_{inf}$ tends to shift
the value of the zero multipole $l_0$ of $C_l^{TE}$ to larger $l$.
For instance, our calculation shows that,
relative to the WMAP-preferred $\beta_{inf}=-2.02$ case,
the exact de Sitter $\beta_{inf}=-2$ case
shifts $l_0$ to a larger value by $\Delta l \sim 1$,
and the less-preferred case $\beta_{inf}=-1.8$
shifts $l_0$ by $\Delta l \sim 8$.

Notice that, for  $C^{TT}_l$ around  $l\sim 20$,
the two curves for $\beta_{inf}=-1.8$ and for $\beta_{inf}=-2.02$
intercept.
A similar interception also occurs for $C^{TE}_l$ as well.
This behavior can be understood as the following.
The initial spectrum of RGWs in Eq. (\ref{initialspectrum})
contains a factor $(\frac{k}{k_0})^{2+\beta_{inf}}$
with the comoving pivot wavenumber $k_0 \simeq 6h^{-1}$.
On large scales $k<k_0$,
one has $(\frac{k}{k_0})^{2+\beta_{inf}}<1$ for $\beta_{inf}=-1.8$,
and  $(\frac{k}{k_0})^{2+\beta_{inf}}>1$ for $\beta_{inf}=-2.02$.
By  the relation in Eq.(\ref{k-l}),
the corresponding pivot multipole is
$l\simeq k\eta_0\simeq 20$.
Thus, in the region of $l<20$, $C^{TT}_l$ and $C^{TE}_l$
have  lower amplitude for $\beta_{inf}=-1.8$,
and  higher amplitude for $\beta_{inf}=-2.02$.

{\large \bf The baryon density  $\Omega_b$ }

The wave equation (\ref{heq}) of RGWs is  not
explicitly coupled with the baryons.
As a result, $h_k(\eta)$ and   $\dot h_k(\eta)$
are not very sensitive to the baryons.
The impact on CMB by the baryons are mainly
through the Thompson scattering terms, $q\xi_k$, $q\beta_k$, and $qG_k$,
   in Boltzmann's equation of photons (\ref{eqxi})  and (\ref{eqbeta}).
During the evolution of CMB,
the photon decoupling process is particularly important,
which depends sensitively on the baryon component.
The fitting formula of the visibility function $V(\eta)$
given in Ref.\cite{Hu95a} contains explicitly the baryon fraction $\Omega_b$.
A larger $\Omega_b$ yields a larger decoupling time $\eta_d$ and
and a smaller decoupling width $\Delta \eta_d$ \cite{Zhao06}.
Moreover, in Eqs. (\ref{ctt2}) through (\ref{cte2}),
the integrands
contain  $h(\eta_d)$,  $\dot{h}(\eta_d)$, and $\Delta\eta_d D(k)$,
which are functions of $\eta_d$ and $\Delta \eta_d$.
We plot  $C^{XX}_l$  for $\Omega_b = 0.045$ and $0.1$ in Fig. \ref{TEBo10}.
It is seen that the amplitudes of $C^{EE}_l$ and $C^{BB}_l$ with
$\Omega_b=0.045$ is slightly higher than those of $\Omega_b=0.1$.
Thus, a larger $\Omega_b$ gives a lower amplitude
of $C^{EE}_l$ and $C^{BB}_l$, agreeing with
the previous calculations \cite{Zhao06}.
As a new result of this paper, Fig. \ref{TEBo10} also shows that
a smaller $\Omega_b$ yields a higher amplitude of $C^{TE}_l$
and shifts the value of the zero multipole $l_0$ to  large $l$.
For instance,
$\Omega_b=0.045$ shifts $l_0$ to a large value by $\Delta l\sim 2$
relative to the $\Omega_b=0.1$ case.
Besides,  $C^{TT}_l$ is less sensitive to
the value of  $\Omega_b$ than the other three spectra.

\begin{center}
{\Large 5. Summary}
\end{center}

In this paper we have presented the approximate, analytical
formulae of the four CMB spectra generated by RGWs.
This has been motivated by an attempt to extract
 signals of RGWs possibly already contained in $C_l^{XX}$,
especially in the magnetic polarization spectrum $C_l^{BB}$
and the cross spectrum $C_l^{TE}$.

In our calculation,
a fitting formula of the exponentially damping factor
$D(k)$, in Eqs.(\ref{DD}) or (\ref{D1}), has been introduced
to describe the decoupling process effectively.
The resulting analytic spectra $C_l^{EE}$ and $C_l^{BB}$
agree quite well with the numerical ones from CAMB
on large scales for the first three peaks for $l\le 600$,
and the error is only $\sim 3\%$.
This improves substantially both the precision and the range of validity
in comparison with the previous analytic studies.
The spectra $C_l^{TT}$ and  $C_l^{TE}$
are first analytically computed in this paper.
They have overall profiles  agreeing with the numerical ones,
but their amplitudes have certain deviations
due to the approximation adopted in {Eq.(\ref{xil}).
More relevant to us is $C_l^{TE}$,
whose amplitude of the $1^{st}$ trough at $l \sim 75$
has a maximum deviation $\sim 20 \%$.
An  analytic formulation
of $\xi(\eta_0)$ better than Eq.(\ref{xil})
should be aimed at in future work.

For the Sachs-Wolfe term in the Boltzmann's equation for photons,
we have included the damping effect of NFS on the RGWs $h_k(\eta)$
as the source.
As is expected,
NFS appreciably reduces the amplitudes of $C_l^{XX}$ for large $l>100$
and, at the same time, shifts slightly
the locations of the peaks to large $l$.
Thus,
in the zero multiple method
by examining the positions where $C^{TE}_l$ crosses $0$,
the shifting due to NFS  effect should be taken into account
for a complete analysis.

We have also demonstrated  the influences on
$C_l^{XX}$ by the tensorial spectrum  index $\beta_{inf}$ of the inflation
and the baryon fraction $\Omega_b$.
It is found that a larger $\beta_{inf}$
leads to a higher amplitude of CMB spectra,
whereas the larger $\Omega_b$ gives a lower one.
Both of them shift the locations of the peaks of $C_l^{XX}$.
In regards to the shifting of the zero multipoles $l_0$ of $C^{TE}_l$,
NFS is  as important as the inflation and the baryons
and should be included in any comprehensive study.

ACKNOWLEDGMENT: T.Y Xia's work is partially
supported by Graduate Student Research Funding from USTC.
Y.Zhang's research work is supported by the CNSF
No.10773009, SRFDP, and CAS.
We thank Dr. W. Zhao and Z. Cai for interesting discussions.


\newpage

\begin{figure}
\caption{
\label{hhdot} The RGWs $h_k(\eta_d)$ and $\dot{h}_k(\eta_d)$ at the
decoupling. NFS reduces the amplitudes and shifts the peaks to
larger $k$. }
\end{figure}

\begin{figure}
\caption{
\label{COMPARE} The analytic spectra $C_l^{XX}$ generated by RGWs are
compared with the numeric ones from CAMB \cite{Lewis}. Here the
decay factor $D(k)$ in Eq.(\ref{DD}) has been used. }
\end{figure}

\begin{figure}
\caption{  \label{BETcomh}
The left panel: the locations of peaks $C^{EE}_l$ and $C^{BB}_l$
correspond to that of  $|\dot{h}_k(\eta_d)|^2$.
The right panel: the locations of peaks of
 $C^{TT}_l$ correspond to that of $|h_k(\eta_d)|^2$.
Here $|\dot{h}_k(\eta_d)|^2$ and $|h_k(\eta_d)|^2$
have been plotted with the variable $k\eta_0$,
which is $\sim l$ by Eq.(\ref{k-l}).}
\end{figure}

\begin{figure}
\caption{  \label{WMAP5}
The predicted $C^{BB}_l$ is well below the constraint of
the 5-year WMAP data \cite{komatsu,nolta}.
Here the tensor/scalar ratio $r=0.37$ is taken in computation.
}
\end{figure}

\begin{figure}
\caption{ \label{TEB}
The NFS modifications on $C_l^{XX}$ are demonstrated.
For $l\le  600$,
the amplitudes are reduced by up to $\sim 35 \%$
and the peaks are shifted to larger $l$ by NFS.
}
\end{figure}

\begin{figure}
\caption{
\label{TEBb18}  $C_l^{XX}$ are sensitive to
the inflation index $\beta_{inf}$ of RGWs.
Two case are plotted for $\beta_{inf}=-1.8$ and $-2.02$.
A larger  $\beta_{inf}$ yields higher amplitudes
of $C_l^{EE}$ and $C_l^{BB}$.
}
\end{figure}
\begin{figure}
\caption{
\label{TEBo10}
The baryon fraction $\Omega_{b}$ affects  $C_l^{XX}$.
Two cases  are given for $\Omega_{b}=0.1$ and $0.045$.
A larger  $\Omega_b$ yields lower amplitudes
of $C_l^{EE}$,  $C_l^{BB}$, and $C_l^{TE}$.
}
\end{figure}

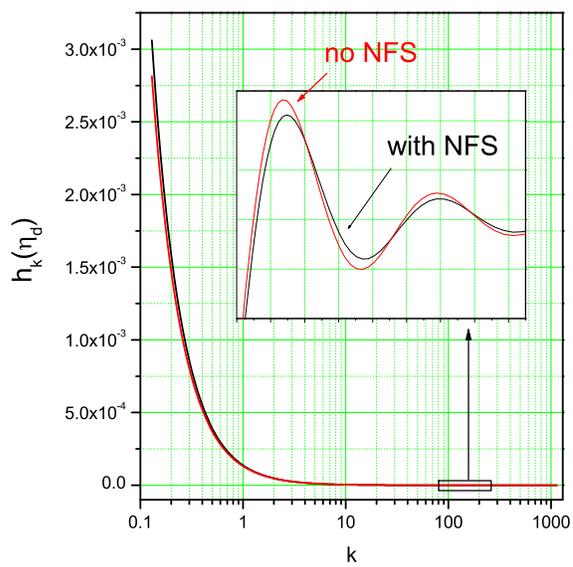
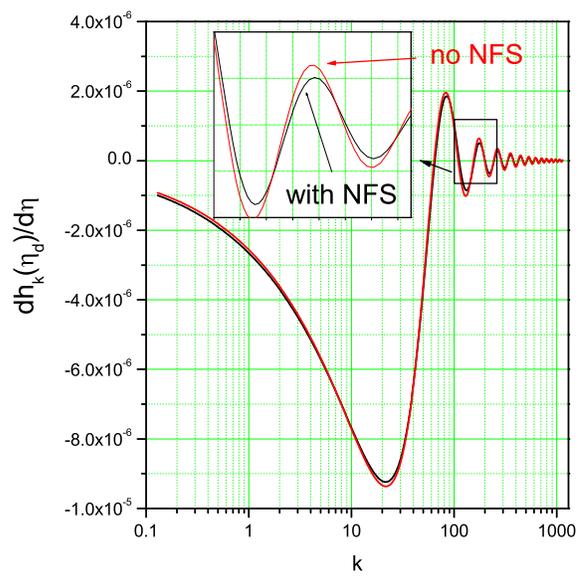
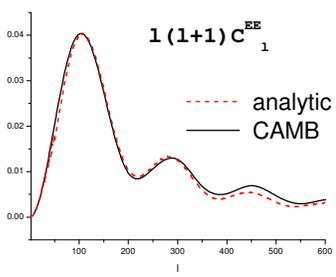
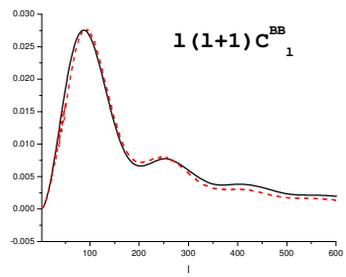
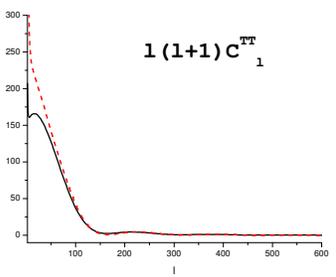
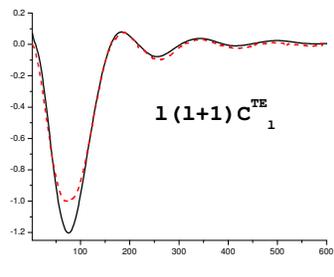
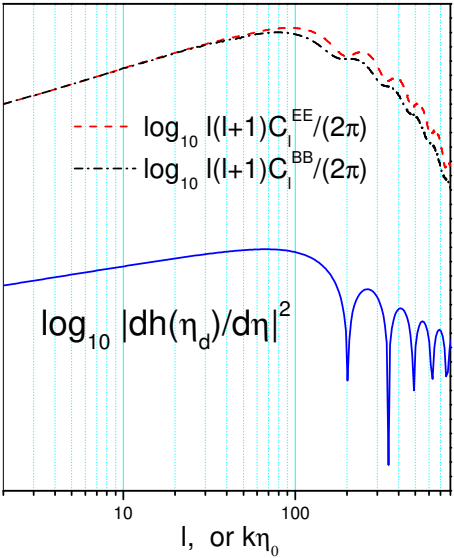
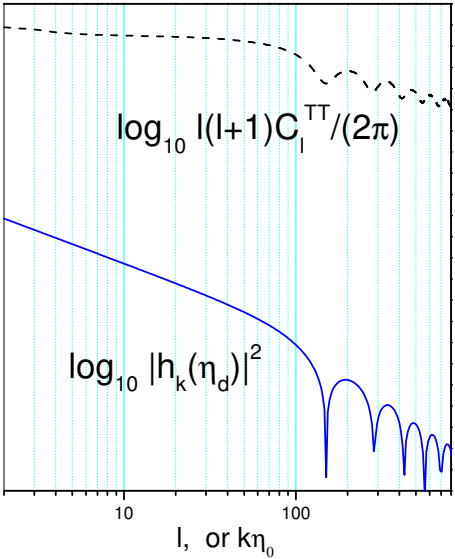
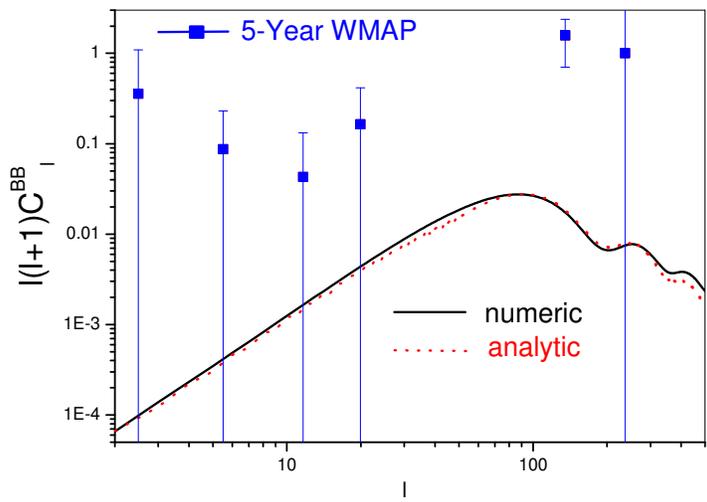
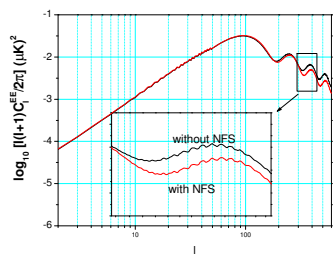
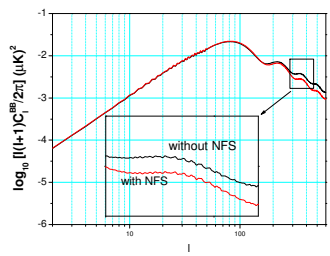
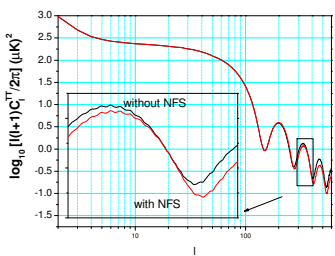
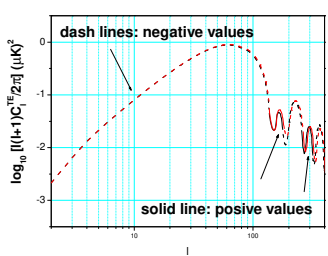
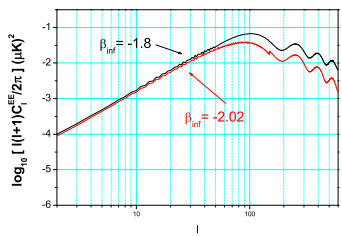
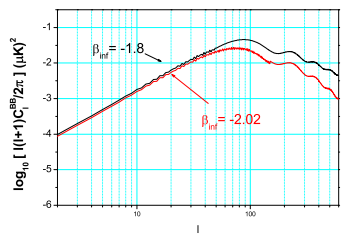
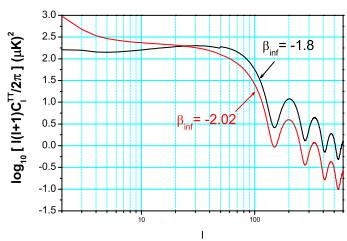
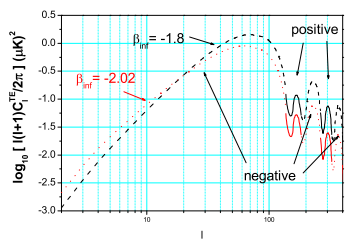
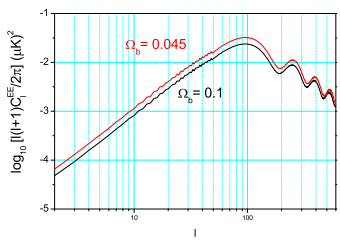
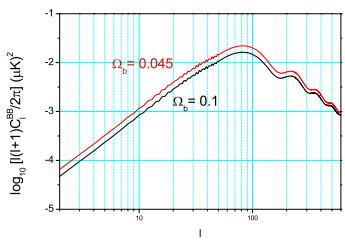
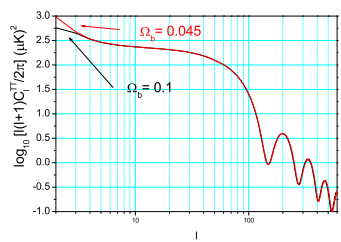
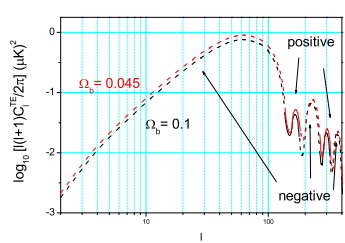
\end{document}